\documentstyle[12pt]{article}
\newcommand{\eeq}{\end{eqnarray}}
\newcommand{\beq}{\begin{eqnarray}}
\newcommand{\bD}{{\bf D}}
\newcommand{\bE}{{\bf E}}
\newcommand{\bB}{{\bf B}}
\newcommand{\bH}{{\bf H}}

\newtheorem{TH}{Theorem}
\date{}
\title{Point Charge in the Born-Infeld Electrodynamics}
\author{\\ \\ Dariusz Chru\'sci\'nski\footnotemark \\
       Fakult\"at f\"ur Physik, Universit\"at Freiburg\\
       Hermann-Herder-Str. 3, D-79104 Freiburg, Germany}

\begin{document}

\def\thefootnote{\relax}\footnotetext{$^*$On leave from
 Institute of Physics, Nicholas Copernicus
University, ul. Grudzi\c{a}dzka 5/7, 87-100 Toru\'n, Poland.}

\maketitle

\begin{abstract}

We show that the nonlinear Born-Infeld field equations supplemented
by the ``dynamical condition'' (certain boundary condition for the
field along the particle's trajectory) define perfectly deterministic
 theory, i.e. particle's trajectory is determined without any equations 
of motion. It is a first step towards constructing the consistent theory
 of point particles interacting with nonlinear electromagnetism.

\end{abstract}

\vspace{4cm}

Freiburg THEP-97/36

\newpage

\section{Introduction}

  Born-Infeld electrodynamics
\cite{BI} was proposed in the thierties as an alternative
for Maxwell theory (see also \cite{IBB} for a useful review).
 Due to the nonlinearity it is very difficult to
solve corresponding field equations (even in the absence of a charged
 matter). Some very specific solutions were found by Pryce 
\cite{Pryce-sol}, however, after the Dirac's paper \cite{Dirac}
on a classical electron and the birth of quantum electrodynamics in
the forties Born-Infeld theory was totally forgoten for a long time.

Recently, there is a new interest in this theory
due to  investigations in  string theory. It turns out that 
some very natural objects in this theory, so called  D-branes,  are
 described by  a kind of nonlinear Born-Infeld
action (see e.g. \cite{brane}). Moreover, due to the remarkable
interest in field and string theory dualities \cite{Olive}, the duality
invariance of Born-Infeld electrodynamics was studied in great details
\cite{duality} (actually this invariance was already observed 
by Schr\"odinger \cite{Schrodinger}).

 In this letter, however, we analyse not a string but 
a classical point charge coupled to  the Born-Infeld nonlinear field.
Why nonlinear electrodynamics? It is well known that Maxwell 
electrodynamics when applied to point-like objects is inconsistent
(see  \cite{Rohrlich} for the review). This
inconsistency originates in the infinite self-energy of the point charge.
In the Born-Infeld theory this self-energy is already finite (actually,
it was Born's motivation to find classical solutions representing 
electrically charged particles with finite self-energy).
Therefore, one may hope that in the theory which gives finite value 
of this quantity it would be possible to describe the particle's
self-interaction in a consistent way. Moreover, the assumption, that
the theory is effectively nonlinear in the vicinity of the charged
particle is very natural from the physical point of view and
this we already  learned from quantum electrodynamics (Born tried
to make contact with quantum field theory by identifying Born-Infeld
Lagrangian as an effective Euler-Heisenberg Lagrangian \cite{Euler}. It
has been shown \cite{Hagiwara1} that the effective Lagrangian can coincide
with those of Born and Infeld up to six-photon interaction terms only).

 We consider here
very specific model of nonlinear theory because, among other nonlinear
theories of electromagnetism, Born-Infeld theory possesses very
distinguished physical properties \cite{Hagiwara2}. For example
it is the only causal spin-1 theory \cite{Plebanski1} (apart from 
the Maxwell one).  Recently, Born-Infeld electrodynamics was
successfully applied \cite{multipole} as a model for generation
of multipole moments of  charged particles.

Our aim in the present letter is to describe the
dynamics of a point charge interacting with Born-Infeld electromagnetic
field. Due to the nonlinearity of the field equations it is impossible
to derive separate equations of motion for the charged particle
corresponding  e.g.  to the celebrated Lorentz-Dirac equation in 
the Maxwell case.  Could we, therefore,
determine the particle's trajectory without equations of motion?
In this letter we show that it is in fact possible. For this purpose
we propose a new approach which was developed in the Maxwell case
in \cite{KIJ}. Analysing the interaction between charged particle
and nonlinear electromagnetism we show that the conservation of
total four-momentum of the composed (particle+field) system is
equivalent to the certain boundary condition for the Born-Infeld
field which has to be satisfied along the particle's trajectory.
We call it ``dynamical condition'' (formula (\ref{dynamical}))
 because, roughly speaking, it replaces particle's equations of motion. 
Field equations supplemented by this condition define perfectly
deterministic theory, i.e. initial data for the particle and field 
uniquely  determine the evolution of the system. 

The same problem was addressed just after the birth of the theory
by Feenberg \cite{Feenberg} and Pryce \cite{Pryce}. They also used the
similar approach, i.e. they considered a conservation law for the 
total energy-momentum tensor.
Therefore, our result
has to be confronted with those obtained 60 years ago.
 In section~6 we show what is the exact relationship between 
three conditions: Feenberg's, Pryce's and ours. It turns out that our
condition given by (\ref{dynamical}) is correct, whereas those of 
Feenberg and Pryce are not consistent (Feenberg's is not consistent with
the field dynamics and Pryce's is not sufficient to determine the particle's
dynamics because it uses not well defined quantities).

Finally, 
we discuss the  physical importance and  relevance of the "dynamical
condition" (\ref{dynamical}) in
constructing  consistent electrodynamics of point-like objects.

\section{Field dynamics}

The Born-Infeld nonlinear electrodynamics \cite{BI}  is based
on the following Lagrangian (we use the Heaviside-Lorentz system of
 units with the velocity of light $c=1$):
\begin{eqnarray}    \label{Lag-BI}
{\cal L}_{BI} &:=&  \sqrt{-\mbox{det}(b\eta_{\mu\nu})}
- \sqrt{-\mbox{det}(b\eta_{\mu\nu} + F_{\mu\nu})}  \nonumber\\
&=& { b^2}\left(1- \sqrt{1-2b^{-2}S - b^{-4}P^2}\right)\ ,
\end{eqnarray}
where $\eta_{\mu\nu}$ denotes the Minkowski metric with the signature
$(-,+,+,+)$ (the theory can be formulated in a general covariant way, 
however, in this paper we will consider only the flat Minkowski space-time).
The standard Lorentz invariants
$S$ and $P$ are defined by:
$S = -\frac 14 F_{\mu\nu}F^{\mu\nu}$ and $P = -\frac 14
F_{\mu\nu}\tilde{F}^{\mu\nu}$ ($\tilde{F}^{\mu\nu}$ denotes the dual tensor).
The arbitrary parameter ``$b$'' has a dimension of a field strength (Born
and Infeld called it the {\it absolute field}) and it measures the
nonlinearity of the theory. In the limit $b \rightarrow \infty$ the
Lagrangian ${\cal L}_{BI}$ tends to the Maxwell Lagrangian $S$.

Adding to (\ref{Lag-BI}) the standard electromagnetic interaction term
``$j^\mu A_\mu$'' we may derive the inhomogeneous field equations
\beq    \label{field-eqs}
\partial_\mu G^{\mu\nu} = - j^\nu\ ,
\eeq
where $G^{\mu\nu} := -2
 {\partial {\cal L}_{BI}}/{\partial F_{\mu\nu}}$. Equations
(\ref{field-eqs}) have formally the same form as Maxwell equations. 
What makes the theory effectively nonlinear are the constitutive
relations, i.e. relations between inductions $(\bD,\bB)$ and
intensities $(\bE,\bH)$:
\begin{eqnarray}
{\bf D}(\bE,\bB) &:=&  \frac{\partial {\cal L}_{BI}}{\partial\bE}
 = \frac{ {\bf E} + b^{-2}({\bf E} {\bf B}){\bf B} }
{\sqrt{ 1 - b^{-2}({\bf E}^2 - {\bf B}^2) - 
b^{-4}({\bf E} {\bf B})^2 }}\ , \label{D} \\ 
{\bf H}(\bE,\bB) &:=& 
-  \frac{\partial {\cal L}_{BI}}{\partial {\bf B}}
= \frac{ {\bf B} - b^{-2}({\bf E} {\bf B}){\bf E} }
{\sqrt{ 1 - b^{-2}({\bf E}^2 - {\bf B}^2) -
 b^{-4}({\bf E} {\bf B})^2 }}\ . \label{H}
\end{eqnarray}
In the Maxwell case we have simply $\bD=\bE$ and $\bH=\bB$.

Fields $\bD$ and $\bB$ play in our analysis important role (they serve
as a Cauchy data for the field evolution). Therefore, it is desirable
to express $\bE$ and $\bH$ in terms of the Cauchy data $\bD$ and $\bB$.
Using (\ref{D}) and (\ref{H}) one easily gets:
\beq
\bE(\bD,\bB) &=& \frac{1}{b^2R} \left[ (b^2 + \bB^2)\bD -
(\bD \bB)\bB \right] \ ,  \label{E}\\
\bH(\bD,\bB) &=& \frac{1}{b^2R} \left[ (b^2 + \bD^2)\bB -
(\bD \bB)\bD \right] \ ,  \label{H1}
\eeq
with $R:= \sqrt{1 + b^{-2}(\bD^2 +\bB^2) +  b^{-4}(\bD\times\bB)^2}$.

Now, let us assume that the external current $j^\mu$ in (\ref{field-eqs})
is produced by a point-like particle moving along the time-like trajectory
 $\zeta$ parameterized by its proper time $\tau$, i.e. $j^\mu(x)=
e\int_{-\infty}^{\infty} d\tau \delta(x-\zeta(\tau))u^\mu(\tau)$. 
This system is very complicated to analyse. In particular, contrary to the
Maxwell case, we do not know the general solution to the inhomogeneous
Born-Infeld field equations. Therefore, following \cite{KIJ}, we
propose the following ``trick''. Instead of solving distribution equations
(\ref{field-eqs}) on the entire Minkowski space-time $\cal M$ let us
treat them as a boundary problem in the region ${\cal M}_\zeta :=
{\cal M} - \{\zeta\}$, i.e. outside the trajectory. In order to well 
pose the problem we have to find an appropriate boundary condition 
which has to be satisfied along $\zeta$, i.e. on the boundary 
${\cal M}_\zeta$.

It turns out that the simplest way to analyse this problem is to use the 
reference frame which is Fermi-transported along $\zeta$ (in this frame the 
particle is always at rest). The general discussion of the accelerated
frame can be found in \cite{MTW}. Obviously, the theory is perfectly
Lorentz invariant, however, the use of this special frame considerably
simplifies our analysis. Let $(y^\mu)$ denotes the standard Lorentz
coordinates in a fixed laboratory frame. At each point $y^\mu(\tau) \in
\zeta$ let $\Sigma_\tau$ denotes the 3-dimensional hyperplane orthogonal 
to $\zeta$. Choose on one $\Sigma_\tau$, say $\Sigma_{\tau_0}$, the system
of cartesian coordinates $(x^k)$, such that the particle is located at
its origin, and transport it ({\it via} the Fermi-transport) to all
other $\Sigma_\tau$. This way we obtain a system $(x^\mu)=(x^0=\tau,x^k)$
of ``co-moving'' coordinates in a neighbourhood of $\zeta$. Obviously,
it is not a global system because different $\Sigma$'s may intersect.
Nevertheless, we will use it globally to describe the evolution of the 
electromagnetic field from one $\Sigma_\tau$ to another (for the 
hyperbolic theory this is well defined problem).

The Born-Infeld field equations have in this frame the following form (see
\cite{KIJ} for the discussion in the Maxwell case):
\beq
\dot{{\bD}} &=& \nabla \times (N\bH)\ , \label{dot-D}\\ 
\dot{{\bB}} &=& -\nabla \times (N\bE)\ , \label{dot-B}
\eeq
where $N=1 + a^kx_k$ (it is the lapse function corresponding to the 
Minkowski metric rewritten in the co-moving frame) and $a^k$ stands for
the rest-frame particle's acceleration. Equations (\ref{dot-D}) and
(\ref{dot-B}) have to be supplemented by the constraints: $\nabla \bD=0$ 
and $\nabla \bB=0$ (note that in $\cal M$ we have 
$\nabla \bD= e\delta_0$).

\section{Asymptotic conditions}

It is well known (\cite{BI} - \cite{IBB}) that $\bE$ and $\bB$ fields are
bounded for $r\rightarrow 0$, whereas $\bD$ and $\bH$ vary as $r^{-2}$
($r$ stands for the radial coordinate, i.e. $r^2 = x^kx_k$). Therefore,
one could think that the standard Lorentz equations of motion can be 
applied in this case. However, despite the fact that $\bE$ and $\bB$ are
bounded, they are not regular in $r=0$ and the Lorentz force
$e(\bE + {\bf v}\times\bB)$ is not well defined.
Let us formally write the following expansions:
\beq
\bE(r) &=& \sum_{n=0}^{\infty} r^n \bE_{(n)}\ , \ \ \ \ \ \ \ \ \ \
\bB(r)  = \sum_{n=0}^{\infty} r^n \bB_{(n)}\ ,\nonumber \\
\bD(r) &=& \sum_{n=-2}^{\infty} r^n \bD_{(n)}\ , \ \ \ \ \ \  \ \  
\bH(r) = \sum_{n=-2}^{\infty} r^n \bH_{(n)}\ , \nonumber
\eeq
where the vectors $\bE_{(n)}$ etc. do not depend on $r$. Let us carefully
analyse the behaviour of the fields for $r\rightarrow 0$ in the co-moving 
frame. In the Maxwell case $\bD^{Maxwell}_{(-2)} = 
e\mbox{\boldmath $r$}/4\pi r$.
 Now, the $\bD_{(-2)}$ term may have much more general form:
\beq   \label{D-2}
\bD_{(-2)} = \frac{e{\cal A}}{4\pi}\frac{\mbox{\boldmath $r$}}{r}\ ,
\eeq
where, due to the Gauss law, the monopole part of the $r$-independent
function $\cal A$ equals 1. Observe, that due to (\ref{dot-D}), $\bH_{(-2)}$
term would have produced an $r^{-3}$ term in $\dot{{\bD}}$, which has to
vanish. Therefore, $\bH_{(-2)} =0$. On the other hand, from (\ref{H1})
it follows that $\bH_{(-2)} =0$ if and only if $\bB_{(0)}=0$. Therefore,
$\bB$ behaves at least like $r$. This information together with 
(\ref{dot-B}) imply the following constraints on $\bE$:
\beq
\nabla N \times \bE_{(0)} + \nabla \times r\bE_{(1)} =0\ .
\eeq
Now, from (\ref{E}) one has $\bE_{(0)}={be}
\mbox{\boldmath $r$}/|e|r$ and, 
therefore
\beq
r\,\nabla \times r\bE_{(1)} = -\frac{be}{|e|}
\, \mbox{\boldmath $a$} \times 
\mbox{\boldmath $r$} \ .
\eeq
The above equation provides the constraint on the transversal part 
$\bE_{(1)}^T$ of $\bE_{(1)}$. Due to $\nabla  r\bE_{(1)}^T =0$,
the transversal part is uniquely given by:
\beq  \label{E-T}
\bE_{(1)}^T = \frac {be}{4|e|}
 \left( 3\mbox{\boldmath $a$} - 
r^{-2}(\mbox{\boldmath $a$} \mbox{\boldmath $r$})
\mbox{\boldmath $r$} \right)\ .
\eeq
This way we have proved the following
\begin{TH}
Any regular solution of Born-Infeld field equations with point-like
external current satisfies (\ref{E-T}).
\end{TH}
Observe, that in the Maxwell case we can derive very similar formula, namely
\beq  \label{E-Max}
\bE_{(-1)} = - \frac{e}{8\pi} \left( \mbox{\boldmath $a$} + 
r^{-2}(\mbox{\boldmath $a$} \mbox{\boldmath $r$})
\mbox{\boldmath $r$} \right)\ .
\eeq
One may easily check that any regular (retarded or advanced) solution 
of Maxwell equations satisfies (\ref{E-Max}) (cf. \cite{KIJ}).

Observe, that (\ref{E-T}), according to our ``boundary philosophy'',
may be interpreted as a boundary condition for $\bE$ on 
$\partial{\cal M}_\zeta$. Due to the hyperbolicity of (\ref{field-eqs})
one may prove
\begin{TH}
The mixed (initial-boundary) value problem for the Born-Infeld equations
in ${\cal M}_\zeta$  with (\ref{E-T}) playing the role of boundary
condition on $\partial{\cal M}_\zeta$ has the unique solution.
\end{TH}

\section{Particle's dynamics}

Up to now our charged particle served only as the point-like external
current for the nonlinear field dynamics. Now, we would like to keep field
and particle's degrees of freedom at the same footing, i.e. we shall
consider a particle as a dynamical object. Of course field equations
alone are not sufficient to uniquely determine the evolution of the
 composed (particle + field) system. Therefore, we impose the conservation 
law of the total four-momentum as the additional equation in the theory.

This point may be further clarified on the level of the boundary
condition (\ref{E-T}). Choosing particle's position {\bf q} and velocity 
{\bf v} as the Cauchy data for the particle's dynamics let us observe
that despite the fact that the time derivatives $(\dot{{\bD}},\dot{{\bB}},
\dot{{\bf q}},\dot{{\bf v}})$ of the Cauchy data are uniquely determined
by the data themselves, the evolution of the composed system is not
uniquely determined. Indeed, $\dot{{\bD}}$ and $\dot{{\bB}}$ are given
by the field equations, $\dot{{\bf q}}={\bf v}$ and $\dot{{\bf v}}$ may
be calculated from (\ref{E-T}). Nevertheless, the initial value problem
is not well posed: keeping the same initial data, particle's trajectory 
can be modified almost at will. This is due to the fact, that now 
(\ref{E-T}) plays no longer the role of boundary condition because we
use it to as a dynamical equation to determine $a^k$. Therefore a new
boundary condition is necessary. We show that this missing condition
is implied by the conservation law of the total four-momentum for
the ``particle + field'' system.

The co-moving components of the total four-momentum are given by:
\beq
 {\cal P}^0(\tau) &=&
 m - \int_{\Sigma_\tau} T^0_{\ 0} d^3x\ ,\label{P-0}\\
{\cal P}_k(\tau) &=&  \int_{\Sigma_\tau}N T^0_{\ k} d^3x\ ,\label{P-k}
\eeq
where $T^\mu_{\ \nu}$ denotes the symmetric energy-momentum tensor
of the Born-Infeld field:
\[ T^\mu_{\ \nu} := \delta^\mu_{\ \nu} {\cal L}_{BI} -
\frac{\partial {\cal L}_{BI}}{\partial S}F^\mu_{\ \lambda}F^\lambda_{\ \nu}
- \frac{\partial {\cal L}_{BI}}{\partial P}
F^\mu_{\ \lambda}\tilde{F}^\lambda_{\ \nu}\ . \]
Using (\ref{Lag-BI}) one easily gets:
\beq
T^{00} &=& b^2 \left(\sqrt{1 + b^{-2}(\bD^2 +\bB^2) +  
b^{-4}(\bD\times\bB)^2} - 1\right)\ ,\nonumber\\
T^{0k} &=& (\bD \times \bB)^k\ ,\nonumber\\
T^{kl} &=& \delta^{kl} \left( \bE\bD + \bH\bB  - T^{00} \right)
- (E^k D^l + H^k B^l ) \ ,\nonumber
\eeq
with $\bE$ and $\bH$ given by (\ref{E}) and (\ref{H1}) respectively.

The factor $N$ in (\ref{P-k}) is necessary because only the co-vector 
$Ndx^0$ and not $dx^0$ is constant on $\Sigma_\tau$ (cf. \cite{MTW}).
This factor is absend in (\ref{P-0}) because the ``upper 0'' introduces
additional $N^{-1}$-factor.
The ``$m$'' in (\ref{P-0}) denotes particle's mass. We stress that we do 
not perform any mass renormalization. The particle's self-energy is finite
and it is already contained in the field energy $\int T^{00}$. Due
to the nonlinearity of the theory there is no way to separate this 
self-energy from the total field energy (this separation is possible
in the Maxwell theory and it enables us to perform mass renormalization,
i.e. to include the infinite self-energy into $m$).

Obviously, ${\cal P}^0$ and ${\cal P}_k$ are not conserved, i.e. they 
depend upon $\tau$. Conserved is the corresponding four-momentum in
the laboratory frame. Therefore, one has to transform ${\cal P}^0$ and 
${\cal P}_k$ to the laboratory frame and compute the time derivatives 
using field equations. However, there is a simpler way to implement
the conservation of momentum in our system. Note, that the corresponding
four-momentum in the laboratory frame is conserved iff ${\cal P}^0$ and 
${\cal P}_k$ are Fermi-transported along $\zeta$ (cf. \cite{MTW}).
Therefore
\beq   \label{dot-P}
\dot{{\cal P}}^0 = - a_k{\cal P}^k\ ,\ \ \ \ \ \ \ 
\dot{{\cal P}}^k = - a^k{\cal P}^0\ .
\eeq
On the other hand, it is easy to show that
\beq    \label{identities}
\partial_\alpha T^\alpha_{\ 0} = Na^k T^0_{\ k}\ , \ \ \ \ \ \ \ 
\partial_\alpha(N T^\alpha_{\ k}) = a_k T^0_{\ 0}\ .
\eeq
In the laboratory frame ($a^k=0$ and $N=1$) these formulae reduce to the
simple conservation law $\partial_\mu T^\mu_{\ \nu}=0$. Now, using
(\ref{identities}), we compute $\dot{{\cal P}}^0$ and $\dot{{\cal P}}^k$: 
\beq   \label{dot-P-0}
\dot{{\cal P}}^0 &=& - \int_{\Sigma^0} \partial_0 T^0_{\ 0}d^3x =
- \lim_{\epsilon\rightarrow 0} \int_{S(\epsilon)} T^\perp_{\ 0} d\sigma -
a^k \int_{\Sigma} NT^0_{\ k} d^3x = - a_k {\cal P}^k \ ,
\eeq
because the surface integral vanishes in the limit $\epsilon\rightarrow 0$
due to the asymptotic condition $\bB_{(0)}=0$. In (\ref{dot-P-0}),
$S(\epsilon)$ denotes 2-sphere with radius $\epsilon$ centered at the particle's 
position, ``$\perp$'' denotes the component perpendicular to $S(\epsilon)$
 and $\Sigma^0 = \Sigma \cap {\cal M}_\zeta \equiv \Sigma - \{0\}$,
where, for simplicity, we skiped the subscript $\tau$. In the same way
\beq   
\dot{{\cal P}}_k &=&  \int_{\Sigma^0} \partial_0 (NT^0_{\ k})d^3x =
 \lim_{\epsilon\rightarrow 0} \int_{S(\epsilon)} NT^\perp_{\ k} d\sigma +
a_k \int_{\Sigma} T^0_{\ 0} d^3x \ .
\eeq
Now, the boundary term does not vanish. Using asymptotic conditions
it is easy to show that
\beq    \label{calka}
 \lim_{\epsilon\rightarrow 0} \int_{S(\epsilon)} NT^\perp_{\ k} d\sigma =
- \frac{|e|b}{4\pi} \int_{S(1)} \frac{x_k}{r}{\cal A}d\sigma =
- \frac{|e|b}{3}{\cal A}_k\ ,
\eeq
where ${\cal A}_k$ is the dipole part of $\cal A$, i.e. DP(${\cal A}) =
:{\cal A}_k x^k/r$. Finally,
\beq   \label{dot-P-k}
\dot{{\cal P}}_k = -a_k {\cal P}^0 - \left( \frac{|e|b}{3}{\cal A}_k 
- ma_k\right)\ .
\eeq
Comparing (\ref{dot-P-k}) with (\ref{dot-P}) we obtain:
\beq   \label{Newton}
ma_k = \frac{|e|b}{3}{\cal A}_k\ .
\eeq
The above equation looks formally like a standard Newton equation. However,
it could not be interpreted as the Newton equation because its r.h.s. is not 
{\it a priori} given (it must be calculated from field equations).

\section{Dynamical condition}

To correctly interpret (\ref{Newton}) we have to take into account
(\ref{E-T}). Now, calculating $a^k$ in terms of $\bE^T_{(1)}$ and 
inserting into (\ref{Newton}) we obtain a relation between $\bE^T_{(1)}$
and $\bD_{(-2)}$. Due to (\ref{E-T}) the radial component 
$(\bE^T_{(1)})^r = ({be}/2|e|r)(a^k x_k)$ and, therefore, $a^k$ equals to the
dipole part of $({2e}/b|e|)(\bE^T_{(1)})^r$. Moreover, from (\ref{D-2}),
DP$(|e|{\cal A})= ({4\pi e}/|e|)\mbox{DP}((\bD_{(-2)})^r)$. 
Therefore, (\ref{E-T}) and
(\ref{Newton}) lead to
\[   \mbox{DP} \left(   \frac{2m}{b} (\bE^T_{(1)})^r -
\frac{4\pi b}{3} (\bD_{(-2)})^r  \right) =0\ .  \]
The above formula may be simplified if we make the following observation:
a charged particle introduces the characteristic length $\lambda_0:=
e^2/6\pi m$
 into the theory. For example, in the Maxwell case $\lambda_0$
appears in the Lorentz-Dirac equation: ${a}^\mu = \frac em 
F^{\mu\nu}_{ext}u_\nu + \lambda_0 (\dot{a}^\mu - a^2 u^\mu)$. The ``$b$''
parameter in the Born-Infeld theory introduces a new scale
 $r_0:= \sqrt{|e|/4\pi b}$. Using $\lambda_0$ and $r_0$ the last formula
may be rewritten as
\beq   \label{dynamical}
   \mbox{DP} \left(   4r_0^4 (\bE^T_{(1)})^r -
\lambda_0 (\bD_{(-2)})^r  \right) =0\ .  
\eeq
Therefore, we finally proved that the conservation of the total  
four-momentum is equivalent to the boundary condition (\ref{dynamical})
for the Born-Infeld field along $\zeta$. We call (\ref{dynamical})
the ``dynamical condition'' for the electrodynamics of a point charge. 
The main result of this letter consists in the following
\begin{TH}
Born-Infeld field equations supplemented by the dynamical condition
(\ref{dynamical}) define perfectly deterministic theory, i.e. initial
data for field and particle uniquely determine the entire evolution
of the system.
\end{TH}

\section{Comparison with  previous results}
\label{Comparison}

Now, we compare (\ref{dynamical}) with the results of \cite{Feenberg} and
\cite{Pryce}. Both authors used the model of a purely electromagnetical
particle (in the ``spirit'' of Einstein's approach to the unitary field 
theory). However, if we put $m=0$ in  (\ref{Newton})
we can compare their results with ours.

Feenberg claimed that the energy and momentum are automatically conserved
due to the field equations and he proposed ``a new dynamical condition
which appears to be singled out from all other possible conditions by
its compelling simplicity". However, his conjecture is not true. It turns
out that the first integral in the r.h.s. of (28) in \cite{Feenberg} does
not vanish (as claimed by Feenberg) but equals exactly to the r.h.s. of
ours (\ref{calka}). The crucial observation in evaluating this integral is
the asymtotic behaviour of the $\bD$ field given by (\ref{D-2}). If one
uses instead of (\ref{D-2}) the Coulomb field  (i.e. ${\cal A}=1$)
this integral vanishes.

Pryce's conclusion is the same as ours, i.e. conservation law for 
energy and momentum imposes the unique condition for the dynamics of 
charged particles. In his approach one has to evaluate the same boundary
integral as Feenberg's (28) and ours (\ref{calka}) (actually this 
integral defines the  force acting on a charge). In \cite{Pryce} it is
given by the first term in the r.h.s. of (5.2). 
Now, to calculate the force he replaced
the point particle by the extended one and obtained very suggestive
(5.12). Obviously, (5.12) defines a force for any extended charge 
distribution.
Pryce claimed that his integral in (5.2) is given by the 
point-particle limit of (5.12). However, it could not be true, 
because (5.12)
is not well defined in this limit: $\bE$ and $\bB$ are not regular at
${\bf x}=0$. Therefore, his dynamical condition is also not well defined.

We evaluated the surface integral in (\ref{calka}) without any use of 
an  extended particle's model. What is crucial for our approach is a
thorough asymptotic analysis of the fields in the vicinity of a charge.
This analysis enables one to calculate (\ref{calka}) using only field
equations outside the particle's trajectory. In our opinion this
``boundary philosophy" is the only consistent way to solve this problem.

\section{Concluding remarks}  \label{Discussion}

Let us now briefly discuss the physical importance of (\ref{dynamical}).
It turns out that the dynamics of the ``particle + field'' system
based on (\ref{dynamical}) may be described by an infinite-dimensional
Hamiltonian system. Both Lagrangian and Hamiltonian formulation of the
above theory will be presented in the next paper. In this letter
we consider only one particle case. However, our result may be generalized
to many particles interacting with nonlinear electromagetism.

At this point the most interesting question arises: is the theory
based on (\ref{dynamical}) 
consistent? We stress that the Theorem~3 does not guarantie the 
consistency of the theory. The analoguous theorem may be proved in 
the Maxwell case \cite{KIJ}, nevertheless, Maxwell electrodynamics
of a point charge is not consistent. To answer this question we
need a precise notion of consistency. There is a very natural definition
of  consistency based on the canonical structure of the theory.
We show in the next paper that according to this definition Born-Infeld
electrodynamics of a point charge is consistent.

It turns out that due to the duality invariance of the Born-Infeld 
electrodynamics \cite{duality} it is possible to describe in the same 
way the dynamics of magnetic monopoles. This problem will be considered 
elsewhere.

There are several open questions. In \cite{Hagiwara2} (see also \cite{Nappi}) 
the Born-Infeld electrodynamics was generalised to the non-abelian 
gauge theories. It would be interesting to apply the approach based on 
(\ref{dynamical}) also in this case.
 Of course one has to ask about the quantum version of this theory. 
This problem  is very difficult. Very little is known about
quantum aspects of the Born-Infeld electrodynamics. Up to our knowledge
only 2 dimensional model was studied in the  sixties \cite{Barbashov}
and recently in \cite{Ikeda}.

\section*{Acknowledgements}

I would like to thank Prof. I. Bia{\l}ynicki-Birula  for 
all  critical but very constructive remarks.
In particular, he suggested  to make the discussion
of the relationship between my result and those of Feenberg and Pryce.
Many thanks are due to Prof. J. Kijowski  and Prof. A. Kossakowski 
for many discussions about the self-interaction
problem in classical electrodynamics. I thank Prof. H. R\"omer 
 for his interest in this work.
Finally, I  thank Alexander von Humboldt Stiftung for 
the financial support.

\end{document}